\documentclass{ws-ijmpe}

\begin{document}

\markboth{Authors' Names}{Instructions for
Typing Manuscripts (Paper's Title)}

\catchline{}{}{}{}{}
%%%%%%%%%%%%%%%%%%%%%%%%%%%%%%%%%%%%%%%%%%%%%%%%%%%%%%%%%%%%%%%%%%%%

\title{CLUSTERING IN RELATIVISTIC DISSOCIATION OF\\
$^9$Be, $^9$C, $^{10}$C AND $^{12}$N NUCLEI
}

%\author{\footnotesize P.I.~ZARUBIN }

\author{D.A.~ARTEMENKOV, V.~BRADNOVA, R.R.~KATTABEKOV, K.Z.~MAMATKULOV\\ N.K.~KORNEGRUTSA, D.O.~KRIVENKOV, A.I.~MALAKHOV, P.A.~RUKOYATKIN\\
V.V.~RUSAKOVA, R.~STANOEVA, I.G.~ZARUBINA, P.I.~ZARUBIN}

\address{Joint Insitute for Nuclear Research\\
Joliot-Curie 6, 141980 Dubna, Moscow region, Russia}

\author{N.G.~PERESADKO, N.G.~POLUKHINA, S.P.~KHARLAMOV}

\address{Lebedev Institute of Physics, Russian Academy of Sciences\\
Leninskij prospekt 53, 119991 Moscow, Russia}

\maketitle

\begin{history}
\received{(received date)}
\revised{(revised date)}
%\accepted{(Day Month Year)}
%\comby{(xxxxxxxxxx)}
\end{history}

\begin{abstract}
The dissociation features in nuclear track emulsion of $^9$Be, $^{9,10}$C, and $^{12}$N nuclei of
 1.2~A~GeV energy are presented. The data presented for the nucleus $^9$Be can be considered as evidence that there
 is a core in its structure in the form of 0$^+$ and 2$^+$ states of the $^8$Be nucleus having roughly equal weights.
 Events of coherent dissociation $^9$C$\rightarrow$3$^3$He associated with the rearrangement of the nucleons outside
 the $\alpha$-clustering are identified. A pattern of the charge fragment topology in the dissociation of
 $^{10}$C and $^{12}$N nuclei is obtained for the first time. Contribution of the unbound nucleus decays to the
 cascade process $^{10}$C$\rightarrow ^9$B$\rightarrow ^8$Be is identified.
\end{abstract}

\section{Introduction}
\indent The concepts of baryonic matter in a cold dilute phase with clustering of nucleons in the lightest
 nuclei $^4$He, $^3$He, $^2$H and $^3$H have been developed in the last decade
 \cite{Schuck07,Typel10,Funaki08,Shlomo09}. Theoretical developments carried out in this
 direction orient towards the study of cluster groups as integral quantum systems and give
 motivation to a new generation of experiments on cluster spectroscopy
 \cite{Oertzen06,Freer07}. Since the macroscopic cluster states can play the role of an
 intermediate phase in astrophysical processes, these studies assume the significance going beyond
 the framework of the nuclear structure problems \cite{Botvina05,Horowitz05}.\par

\indent The method of nuclear track emulsion provides a uniquely complete observation of multiple
 fragment systems produced in dissociation of relativistic nuclei. Approximate conservation of the
 initial momentum per nucleon by relativistic fragments is used in the kinematical analysis of the
 events to compensate the lack of momentum measurements. The fragmenting system excitation can be 
 defined as Q = M$^*$-M, where M$^*$ is the invariant mass and � � the projectile mass or total
 fragment mass. The value M$^*$ is defined by the relation
 M$^{*2}$ = ($\sum$P$_j$)$^2$=$\sum$(P$_i \cdot$P$_k$), where P$_{i,k}$ 4-momenta fragments
 i and the k, determined in the approximation of the conservation of the primary momentum value per
 nucleon.\par
 
\indent The most valuable events of coherent dissociation of nuclei in narrow jets of light and the
 lightest nuclei with a net charge as in the initial nucleus, occurring without the formation of
 fragments of the target nuclei and mesons (the so-called \lq\lq white\rq\rq stars)
 \cite{Andreeva05,Andreeva06}, comprise a few percent among the observed interactions. The data on
 this phenomenon are fragmented, and the interpretation is not offered. The dissociation degree of
 light O, Ne, Mg and Si, and as well as heavy Au, Pb and U nuclei may reach a complete destruction
 to light and the lightest nuclei and nucleons, resulting in cluster systems of an unprecedented
 complexity. The dissociation dynamics of heavy nuclei can be grounded on dissociation peculiarities
 established for light nuclei. An extensive collection of photographs of such interactions is
 gathered by the BECQUEREL collaboration \cite{web}.\par
 
\indent Despite the fact that the potential of the relativistic approach to the study of nuclear
 clustering is recognized long ago, e-experiments were not be able to get closer to the required 
 detailed observation of the relativistic fragment ensembles. The related pause has led to the 
 proposal to irradiate nuclear track emulsion in the JINR Nuclotron beams of the whole family of
 1.2~A~GeV light nuclei, including radioactive ones \cite{Artemenkov071,Shchedrina07}. Studies
 with relativistic neutron-deficient nuclei have special advantages due to more complete
 observations. The dissociation features of $^9$Be, $^{9,10}$C, and $^{12}$N nuclei, which are
 the sources of basic cluster configurations, will be presented in the present paper.\par

\section{Dissociation $^9$Be$\rightarrow$2$\alpha$}
\indent In order to justify the application of the relativistic fragmentation in the study of N$\alpha$-systems
 \cite{Andreeva06} it was suggested to investigate the dynamics of the formation of $\alpha$-particle pairs at a
 high statistical level and under the simplest conditions (without combinatorial background) which are provided in
 the relativistic fragmentation $^9$Be$\rightarrow$2$\alpha$ \cite{Artemenkov071,Artemenkov072,Artemenkov08}.
 The secondary $^9$Be beam was obtained by fragmentation of accelerated $^{10}$B nuclei
 \cite{Rukoyatkin08}. When scanning the exposed emulsion 500 events $^9$Be$\rightarrow$2$\alpha$ in a
 fragmentation cone of 0.1 rad have been found. About 81\% $\alpha$-pairs form roughly equal groups on
 $\Theta_{2\alpha}$: \lq\lq narrow\rq\rq (0 $< \Theta_{n} <$ 10.5 mrad) and
 \lq\lq wide\rq\rq (15.0 $< \Theta_{w} <$ 45.0 mrad) ones. The $\Theta_{n}$ pairs are consistent with $^8$Be
 decays from the ground state 0$^+$, and pairs $\Theta_{w}$ - from the first excited state 2$^+$. The
 $\Theta_{n}$ and $\Theta_{w}$ fractions are equal to 0.56$\pm$0.04 and 0.44$\pm$0.04. These values are well
 corresponding to the weights of the $^8$Be 0$^+$ and 2$^+$ states $\omega_{0^+}$ = 0.54 and $\omega_{2^+}$ = 0.47
 in the two-body model n - $^8$Be, used to calculate the magnetic moment of the $^9$Be nucleus 
 \cite{Parfenova051,Parfenova052}.\par
 
\indent For the coherent dissociation $^9$Be$\rightarrow$2$\alpha$+n, the average value of the total $\alpha$-pair
 transverse momentum is equal to $<P_{Tsum}> \approx$ 80~MeV/c, which can be assigned to the average transverse
 momentum carried away by neutrons. For the $^9$Be coherent dissociation through the $^8$Be 0$^+$ and 2$^+$ states there
 is no differences in the values $<P_{Tsum}>$, which points to a \lq\lq cold fragmentation\rq\rq mechanism. The
 whole complex of these observations may serve as an evidence of the simultaneous presence of the $^8$Be 0$^+$ and
 2$^+$ states with similar weights in the ground state of the nucleus $^9$Be.\par

\section{\label{sec:level3}Coherent dissociation of $^9$C nuclei}
\indent One can expect that the pattern established for the $^7$Be \cite{Peresadko07} and $^8$B
 \cite{Stanoeva07,Stanoeva09} nuclei is reproduced for nucleus $^9$C with the addition of one or two protons.
 In addition, the emergence of a 3$^3$He ensemble becomes possible. An intriguing hypothesis is that in the
 nuclear astrophysical processes the 3$^3$He system can be a 3$\alpha$-process analog.\par

\indent A secondary beam, optimized for $^9$C nucleus selection was formed by fragmentation of accelerated
 $^{12}$C nuclei \cite{Krivenkov10,Rukoyatkin08}. It was important in this irradiation to avoid overexposure by
 the accompanying flux of $^3$He nuclei. The intensity ratio of the nuclei with charges Z$_{pr}$ = 6 and 2
 amounted to 1 : 10. This factor has limited statistics and made the scan for $^9$C interactions much more labor 
 demanding.\par
\indent Among the total number of \lq\lq white\rq\rq stars N$_{ws}$, detected in this exposure, 15 events
 $^9$C$\rightarrow ^8$B+p and 16 events $^7$Be+2p are found. Statistics in the channels 2He+2H (24), He+4H (28)
 and 6H (6) well corresponds to the $^7$Be core dissociation. The event fraction $^9$C$\rightarrow$3$^3$He (16)
 was found to be the same as that of the channels $^9$C$\rightarrow ^8$B+p and $^7$Be+2p.\par

 \indent The latter fact can point to a significant admixture of a virtual 3$^3$He state in the $^9$C ground
 state. This component can give a contribution to the $^9$C magnetic moment, which has an abnormal value in terms
 of the shell model \cite{Utsuno04}.\par

\section{\label{sec:level4}Coherent dissociation of $^{10}$C and $^{12}$N nuclei}
\indent The $^{10}$C nucleus is the only example of the system, which has the \lq\lq super-boromean\rq\rq 
properties, since the removal of one of the four clusters in the 2$\alpha$+2p structure (threshold 3.8~MeV)
 leads to an unbound state. The particular feature of the $^{12}$N nucleus consists in the low proton separation 
 threshold (600~keV). Furthermore, the dissociation can occur through the channels $\alpha$+$^8$B (8~MeV)
 , p+$^7$Be+$\alpha$, as well as into more complicated ensembles with the $^7$Be core break.\par
\indent Generation of $^{12}$N and $^{10}$C nuclei is possible in charge exchange and fragmentation reactions
 of accelerated $^{12}$C nuclei \cite{Rukoyatkin08}. The charge to weight ratio Z$_{pr}$/A$_{pr}$ differs by 
 only 3\% for these nuclei, while the momentum acceptance of the separating channel is 2 - 3\%. Therefore, their
 separation is not possible, and the $^{12}$N and $^{10}$C nuclei are simultaneously present in the secondary beam,
 forming a so-called beam \lq\lq cocktail\rq\rq. The contribution of $^{12}$N nuclei is small in respect to
 $^{10}$C ones in accordance with the cross sections for charge transfer and fragmentation reactions. Also,
 the beam contains $^7$Be nuclei, differing by Z$_{pr}$/A$_{pr}$ from $^{12}$N nuclei only by 2\%. \par
\indent Due to the momentum spread $^3$He nuclei can penetrate in the separating channel. For neighboring
 $^8$B, $^9$C and $^{11}$C nuclei the difference by Z$_{pr}$/A$_{pr}$ from $^{12}$N is about 10\%, which leads
 to suppression in these isotopes. Identification of $^{12}$N nuclei can be performed by $\delta$-electron
 counting along the beam tracks. In the $^{10}$C case, relying on the charge topology of the produced
 \lq\lq white\rq\rq stars it is necessary to be sure that the neighboring carbon isotope contribution is small.
 These considerations provided the justification to expose nuclear track emulsion in a mixed beam of $^{12}$N,
 $^{10}$C and $^7$Be nuclei.\par
\indent The initial scanning phase consisted in visual search of beam tracks with charges Z$_{pr}$=1, 2 and
 Z$_{pr}>$2. The ratio of beam tracks with charges Z$_{pr}$= 1, 2 and Z$_{pr} >$2 is found to be equal
 $\approx$ 1 : 3 : 18. Thus, the contribution of $^3$He nuclei dramatically decreased compared with the $^9$C
 irradiation, which radically raised the event search efficiency.\par
\indent The presence of fragments Z$_{fr}>$2 makes the charge identification of beam Z$_{pr}$ and secondary
 Z$_{fr}$ tracks necessary. For the calibration the average density of $\delta$-electrons N$_{\delta}$ was 
 measured along the beam tracks, which produced the \lq\lq white\rq\rq stars 2He+2H, 2He and He+2H, and also stars
 with fragments Z$_{fr}>$2 ($^{12}$N candidates). Thus, the correlation between the charge topology $\sum$Z$_{fr}$
 and N$_{\delta}$ was established which permitted to determine beam track charges Z$_{pr}$ and the fragment
 charges Z$_{fr}>$2.\par
\indent For \lq\lq white\rq\rq stars N$_{ws}$ with charge topology $\sum$Z$_{fr}$=6 the most probable channel is
 represented by 91 events 2He+2H, which might be expected for the isotope $^{10}$C. The channel He+4H is found to
 be suppressed (14 events), as in the $^{10}$C case it is required to overcome the high threshold of the
 $\alpha$-cluster break up.\par
\indent In this irradiation 20 \lq\lq white\rq\rq stars with Z$_{pr}$=7 and $\sum$Z$_{fr}$=7 are found,
 corresponding to the dissociation of $^{12}$N nuclei.  There are the following channels among them: C+H (1),
 $^7$Be+He+H (2), $^7$Be+3H (4), $^8$B+2H (3), 3He+H (2), 2He+3H (6), He+5H (3). Thus, half of the events contain 
 a fragment Z$_{fr}>$ 2, clearly differing from the cases of nuclei $^{14}$N \cite{Artemenkov071,Shchedrina07} 
 and $^{10}$C.\par

\section{\label{sec:level5}Production of $^8$Be and $^9$B nuclei in $^{10}$C dissociation }   
\indent  The unbound $^8$Be nucleus plays the role of the core in the $^{10}$C structure, which should be
 manifested in the fragmentation intensity $^{10}$C$\rightarrow ^8$Be.  Distribution $\alpha$-pairs in
 the 91 \lq\lq white\rq\rq stars 2He+2H on the excitation energy Q$_{2\alpha}$ is presented in Fig. 1. In 30 
 events the Q$_{2\alpha}$ value does not exceed 500 keV (inset in Fig. 1). For them, the average value is
 $<Q_{2\alpha}>\approx$110$\pm$20~keV and the mean-square scattering $\sigma$=40~keV, which well corresponds 
 to the decays of the $^8$Be 0$^+$ ground state. The $^8$Be 0$^+$ contribution is approximately the same as for
 the neighboring cluster nuclei.\par
\indent The unbound $^9$B nucleus can be another major product of the $^{10}$C coherent dissociation. Fig. 2 shows
 the distribution of \lq\lq white\rq\rq stars 2He+2H on the excitation energy Q$_{2\alpha p}$, defined by the
 difference of the invariant mass of the three fragments 2$\alpha$+p and the mass of the proton and the doubled
 $\alpha$-particle mass. The Q$_{2\alpha p}$ values for one of two possible 2$\alpha$+p triples do not exceed
 500~keV in 27 events (inset in Fig. 2). The average value for these triples is $<Q_{2\alpha p}>$=250$\pm$15~keV
 with rms $\sigma$=74~keV. These values well correspond to the $^9$B ground state decay via the channel p+$^8$Be
 (0$^+$) having the published values of energy 185~keV and width (0.54$\pm$0.21)~keV\cite{tunl}. In the region 
 limited by Q$_{2\alpha}<$1~MeV ane time both triples correspond to the decay of the nucleus $^9$B. In all
 other $^9$B cases one of  Q$_{2\alpha p}$ is above 500 keV.\par
\indent In addition, excitations $\alpha$+2p are studied on the remaining statistics of \lq\lq white\rq\rq stars
 2He+2H beyond $^9$B decays. In the spectrum of Q$_{2\alpha p}$, there is no clear signal of $^6$Be decays
 \cite{web}, and its estimated contribution does not exceed 20\%. This aspect deserves further analysis taking
 the proton angular correlations into account.\par

\begin{figure}[th]
\centerline{\psfig{file=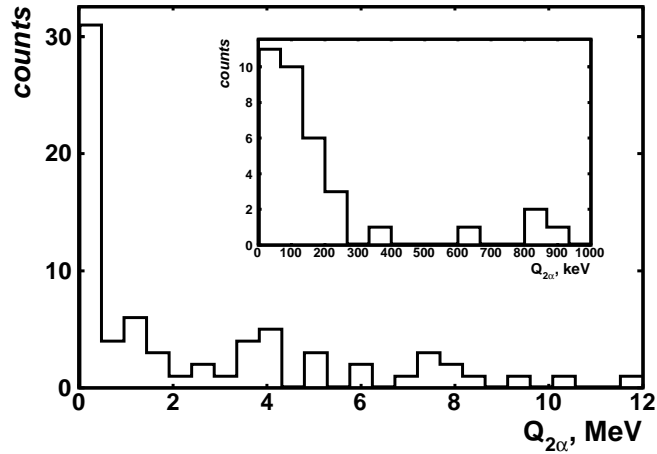, width=10cm}}
\vspace*{8pt}
\caption{Distribution number of \lq\lq white\rq\rq stars N$_{ws}$ of the 2He+2H topology on the excitation energy Q$_{2\alpha}$ pairs of $\alpha$-particles in the inset - enlarged distribution Q$_{2\alpha}$}
\end{figure}

\begin{figure}[th]
\centerline{\psfig{file=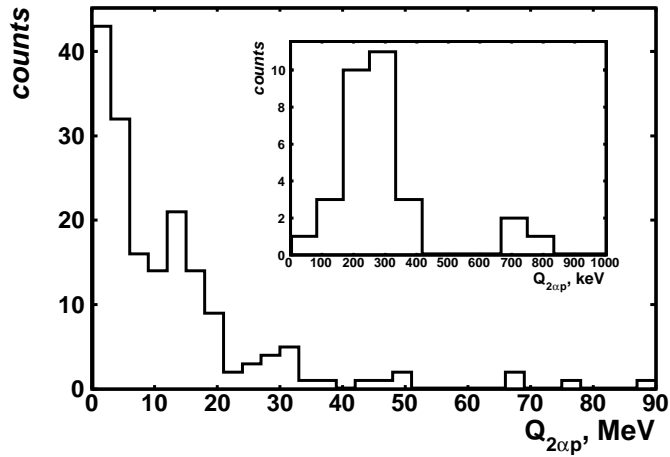, width=10cm}}
\vspace*{8pt}
\caption{Distribution number of \lq\lq white\rq\rq stars N$_{ws}$ of the 2He+2H topology on the excitation energy Q$_{2\alpha p}$ triples 2$\alpha$+p; in the inset - enlarged distribution Q$_{2\alpha p}$}
\end{figure}

\section*{Acknowledgements}

\indent The work was supported by grants of the Russian Foundation for Basic Research
 (96-1596423, 02-02-164-12�, 03-02-16134, 03-02-17079, 04-02-17151. 04-02-16593 and
09-02-9126 CT-a) and grants of
 the JINR
 Plenipotentiaries of the Republic of Bulgaria, the Slovak Republic, the Czech Republic
 and Romania in the years 2002-2010.\par

\end{document}